# Transient Error Analysis of the LMS and RLS Algorithm for Graph Signal Estimation

Haiquan Zhao, *Senior Member*, *IEEE* and Chengjin Li

*Abstract*—Recently, the proposal of the least mean square (LMS) and recursive least squares (RLS) algorithm for graph signal processing (GSP) provides excellent solutions for processing signals defined on irregular structures such as sensor networks. The existing work has completed the steady state error analysis of the GSP LMS algorithm and GSP RLS algorithm in Gaussian noise scenarios, and a range of values for the step size of the GSP LMS algorithm has also been given. Meanwhile, the transient error analysis of the GSP LMS algorithm and GSP RLS algorithm is also important and challenging. Completing the above work will help to quantitatively analyze the performance of the graph signal adaptive estimation algorithm at transient moments, which is what this paper is working on. By using formula derivation and mathematical induction, the transient errors expressions of the GSP LMS and GSP RLS algorithm are given in this paper. Based on the Brazilian temperature dataset, the related simulation experiments are executed, which strongly demonstrate the correctness of our proposed theoretical analysis.

## I. INTRODUCTION

Recently, with the rapid development of Internet of Things (IoT) technology, sensor technology [1], [2], [3], the demand for signal processing defined on irregular structures is becoming more and more vigorous. However, conventional adaptive filtering techniques cannot effectively handle the above types of signals, which lies in the fact that the theoretical foundations and processing means of the former conflict with brand new concepts such as graph structure. Thankfully, the introduction of graph theory has effectively resolved this paradox. Based on graph theory, the graph signal processing (GSP) is proposed, which involves concepts such as the definition of a graph signal, frequency domain representation [4], [5], bandwidth characteristics [6], [7], [8], and graph sampling strategy [9], [10].

By combining the well-established adaptive filtering field with graph signal, developing of a family of algorithms for adaptive estimation of graph signal is attracting more and more attention. Inspired by the conventional least mean square (LMS) strategy, the GSP LMS algorithm is proposed, which is seen as pioneering a new field of signal processing [11]. The GSP LMS algorithm is simple and has a low computational complexity, but with a slow convergence rate. To solve this problem, based on the idea of traditional recursive least squares (RLS) algorithm, the GSP RLS algorithm is proposed, which has much faster convergence rate but its accompanying cost is higher computational burden [12], [13]. By performing detailed and rigorous mathematical analysis, the theoretical steady state errors analysis of the GSP LMS and GSP RLS algorithm in Gaussian noise scenarios is given in [11], which has been strongly verified by simulation experiments.

In addition to the mentioned above work, it is also important to analyze the transient error behavior of the GSP LMS and GSP RLS algorithm in Gaussian noise scenarios. Compared to the existing work on transient error analysis of conventional adaptive filtering algorithms [14], [15], [16], [17], [18], there are many different concepts in graph signal adaptive estimation algorithms. Compared to traditional adaptive filtering algorithms, the graph signal adaptive estimation algorithm broadens the signal processing scenario to topological graphs with multiple nodes, which involve brand new concepts such as graph structure matrix, graph sampling matrix, spectrally sparse parameter $F$ and so on [19], [20], [21], [22], [23]. Thus, it is necessary to analyze the transient error behavior of the GSP LMS and GSP RLS algorithm in conjunction with brand new concepts such as graph structure matrix.

In this paper, by using formula derivation and mathematical induction, the theoretical transient errors analysis of the GSP LMS and GSP RLS algorithm is given. The related simulation experiments are performed based on the Brazilian temperature dataset, and the simulation results validate the correctness of our proposed theoretical analysis.

## II. REVIEW OF THE RELATED ALGORITHMS

When it comes to a graph signal adaptive estimation algorithm, the noisy graph signal $\mathbf{x}_w(t) \in \mathbb{R}^{N \times 1}$ and the error signal $\mathbf{e}(t) \in \mathbb{R}^{N \times 1}$ at discrete moment $t$ are shown as follows.

$$\mathbf{x}_w(t) = \mathbf{x}_o + \mathbf{w}(t) = \mathbf{U}_F \mathbf{s}_F + \mathbf{w}(t) \qquad (1)$$

$$\begin{aligned}\mathbf{e}(t) &= \mathbf{D}_\mathbf{S}\left(\mathbf{x}_w(t) - \mathbf{U}_F \hat{\mathbf{s}}_F(t)\right) \\ &= \mathbf{D}_\mathbf{S}\left(\mathbf{x}_w(t) - \hat{\mathbf{x}}_o(t)\right)\end{aligned} \qquad (2)$$

where $\mathbf{x}_o \in \mathbb{R}^{N \times 1}$ is the actual graph signal, $\mathbf{w}(t) \in \mathbb{R}^{N \times 1}$ is the noise signal, which satisfies $\mathrm{E}\left[\mathbf{w}(t)\mathbf{w}^T(t)\right] = \mathbf{C}_w(t)$. $\mathbf{U}_F \in \mathbb{R}^{N \times F}$ represents the graph structure, $F$ is the spectrally sparse parameter, once $F$ is determined, $\mathbf{U}_F$ will be deter-

This work was in part by National Natural Science Foundation of China (grant: 62171388, 61871461, 61571374) and the Fundamental Research Funds for the Central Universities(2682025GH026)

Haiquan Zhao, Chengjin Li are with the Key Laboratory of Magnetic Suspension Technology and Maglev Vehicle, Ministry of Education, and the School of Electrical Engineering, Southwest Jiaotong University, Chengdu, 610031, China (e-mail: hqzhao_swjtu@126.com, chengjinliswjtu@126.com)

Corresponding author: Haiquan Zhao

mined with it [10], [11], [19]. $\mathbf{s}_F \in \mathbb{R}^{F\times 1}$ is the frequency domain representation of $\mathbf{x}_o$. After $\mathbf{U}_F$ is determined, $\mathbf{s}_F$ will be determined accordingly. $\hat{\mathbf{x}}_o(t)$ is the estimation of $\mathbf{x}_o$ at discrete moment $t$, $\hat{\mathbf{s}}_F(t) \in \mathbb{R}^{F\times 1}$ is the frequency domain representation of $\hat{\mathbf{x}}_o(t)$ [6], [7]. $\mathbf{D_S}$ is the graph sampling matrix, a diagonal matrix whose element on the diagonal is only 0 or 1, which depends on the sampling set $|\mathbf{S}|$ [8], [9], [10].

The update formula of the GSP LMS algorithm is as follows.

$$\hat{\mathbf{x}}_o(t+1) = \hat{\mathbf{x}}_o(t) + \mu \mathbf{U}_F \mathbf{U}_F^T \mathbf{e}(t) \tag{3}$$

When it comes to the GSP RLS algorithm, its graph signal update formula is shown in (4).

$$\hat{\mathbf{x}}_o(t+1) = \hat{\mathbf{x}}_o(t) + (1-\lambda)\mathbf{U}_F \mathbf{M}' \mathbf{U}_F^T \mathbf{D_S} \mathbf{C}_w^{-1}(t)\mathbf{e}(t) \tag{4}$$

where $\lambda$ is the forgetting factor, which satisfies $0 \ll \lambda \leq 1$, and $\mathbf{M}' = \left(\mathbf{U}_F^T \mathbf{D_S} \mathbf{C}_w^{-1} \mathbf{D_S} \mathbf{U}_F\right)^{-1}$.

The metric $\mathrm{MSD}_G(t)$, which is used to measure the performance of a graph signal adaptive estimation algorithm, is given as follows

$$\begin{aligned}\mathrm{MSD}_G(t) &= \left\|\hat{\mathbf{x}}_o(t) - \mathbf{x}_o\right\|_2^2 \\ &= \left\|\mathbf{U}_F \hat{\mathbf{s}}_F(t) - \mathbf{U}_F \mathbf{s}_F\right\|_2^2 \\ &= \left\|\mathbf{U}_F \Delta\hat{\mathbf{s}}_F(t)\right\|_2^2 \\ &= \Delta\hat{\mathbf{s}}_F^T(t) \mathbf{U}_F^T \mathbf{U}_F \Delta\hat{\mathbf{s}}_F(t)\end{aligned} \tag{5}$$

Since $\mathbf{U}_F^T \mathbf{U}_F = \mathbf{I}$, it can be obtained that

$$\mathrm{MSD}_G(t) = \Delta\hat{\mathbf{s}}_F^T(t) \Delta\hat{\mathbf{s}}_F(t) \tag{6}$$

III. TRANSIENT ERROR ANALYSIS OF THE GSP LMS ALGORITHM AND GSP RLS ALGORITHM

For ease of processing, the following reasonable assumptions are adopted for performing the related transient error analysis.

A1    $\mathbf{x}$ is bandlimited or is sparse in the graph structure.
A2    $\mathbf{U}_F$, $\Delta\hat{\mathbf{s}}_F(t)$ and $\mathbf{s}_F$ are independent of each other.
A3    The $\mathbf{w}(t)$ is the zero-mean Gaussian noise which satisfies $\mathrm{E}\left[\mathbf{w}(t)\mathbf{w}^T(t)\right] = \mathbf{C}_w(t)$, $\mathbf{C}_w(t) = \mathbf{C}_w$
A4    $F$ is taken to be a fixed value, which leads that $\mathbf{U}_F$ a constant matrix.
A5    $|\mathbf{S}|$ is taken to be a fixed value, which leads that $\mathbf{D_S}$ a constant matrix.

*A. Transient Error Analysis of the GSP LMS Algorithm*

Based on (3), $\Delta\hat{\mathbf{s}}_F(t) = \hat{\mathbf{s}}_F(t) - \mathbf{s}_F$, it can be obtained that

$$\Delta\hat{\mathbf{s}}_F(t+1) = \left(\mathbf{I} - \mu\mathbf{U}_F^T \mathbf{D_S} \mathbf{U}_F\right)\Delta\hat{\mathbf{s}}_F(t) + \mu\mathbf{U}_F^T \mathbf{D_S} \mathbf{w}(t) \tag{7}$$

$\Delta\hat{\mathbf{s}}_F(1)$ is defined as follows

$$\begin{aligned}\Delta\hat{\mathbf{s}}_F(1) &= \hat{\mathbf{s}}_F(t) - \mathbf{s}_F \\ &= \mathbf{0}_{F\times 1} - \mathbf{s}_F \\ &= -\mathbf{s}_F\end{aligned} \tag{8}$$

Let $\mathbf{I} - \mu\mathbf{U}_F^T \mathbf{D_S} \mathbf{U}_F = \mathbf{A}$, according to (7) and (8), we can obtain $\Delta\hat{\mathbf{s}}_F(2)$, which is as follows

$$\begin{aligned}\Delta\hat{\mathbf{s}}_F(2) &= \mathbf{A}\Delta\hat{\mathbf{s}}_F(1) + \mu\mathbf{U}_F^T \mathbf{D_S} \mathbf{w}(t) \\ &= -\mathbf{A}\mathbf{s}_F + \mu\mathbf{U}_F^T \mathbf{D_S} \mathbf{w}(t)\end{aligned} \tag{9}$$

$\Delta\hat{\mathbf{s}}_F(3)$ is obtained as follows:

$$\begin{aligned}\Delta\hat{\mathbf{s}}_F(3) &= \mathbf{A}\Delta\hat{\mathbf{s}}_F(2) + \mu\mathbf{U}_F^T \mathbf{D_S} \mathbf{w}(t) \\ &= \mathbf{A}\left(-\mathbf{A}\mathbf{s}_F + \mu\mathbf{U}_F^T \mathbf{D_S} \mathbf{w}(t)\right) + \mu\mathbf{U}_F^T \mathbf{D_S} \mathbf{w}(t) \\ &= -\mathbf{A}^2\mathbf{s}_F + \mu\mathbf{A}\mathbf{U}_F^T \mathbf{D_S} \mathbf{w}(t) + \mu\mathbf{U}_F^T \mathbf{D_S} \mathbf{w}(t)\end{aligned} \tag{10}$$

According to (8), (9) and (10), combined with mathematical induction, it can be concluded that

$$\begin{aligned}\Delta\hat{\mathbf{s}}_F(t) = &-\mathbf{A}^{t-1}\mathbf{s}_F + \mathbf{A}^{t-2}\mu\mathbf{U}_F^T \mathbf{D_S} \mathbf{w}(t) + \mathbf{A}^{t-3}\mu\mathbf{U}_F^T \mathbf{D_S} \mathbf{w}(t) + \ldots \\ &+ \mu\mathbf{U}_F^T \mathbf{D_S} \mathbf{w}(t)\end{aligned} \tag{11}$$

The expression of $\mathbf{K}$ is defined in (12), by multiplying $\mathbf{A}$ on the left-hand side of $\mathbf{K}$, it can be obtained that

$$\mathbf{K} = \mathbf{A}^{t-2}\mu\mathbf{U}_F^T \mathbf{D_S} \mathbf{w}(t) + \mathbf{A}^{t-3}\mu\mathbf{U}_F^T \mathbf{D_S} \mathbf{w}(t) + \ldots + \mu\mathbf{U}_F^T \mathbf{D_S} \mathbf{w}(t) \tag{12}$$

$$\mathbf{AK} = \mathbf{A}^{t-1}\mu\mathbf{U}_F^T \mathbf{D_S} \mathbf{w}(t) + \mathbf{A}^{t-2}\mu\mathbf{U}_F^T \mathbf{D_S} \mathbf{w}(t) + \ldots + \mu\mathbf{A}\mathbf{U}_F^T \mathbf{D_S} \mathbf{w}(t) \tag{13}$$

Subtracting (13) from (12), we can obtain

$$(\mathbf{A} - \mathbf{I})\mathbf{K} = \left(\mathbf{A}^{t-1} - \mathbf{I}\right)\mu\mathbf{U}_F^T \mathbf{D_S} \mathbf{w}(t) \tag{14}$$

Thus, the expression of $\mathbf{K}$ is as follows:

$$\begin{aligned}\mathbf{K} &= (\mathbf{A} - \mathbf{I})^{-1}\left(\mathbf{A}^{t-1} - \mathbf{I}\right)\mu\mathbf{U}_F^T \mathbf{D_S} \mathbf{w}(t) \\ &= \left(\mathbf{I} - \mu\mathbf{U}_F^T \mathbf{D_S} \mathbf{U}_F - \mathbf{I}\right)^{-1}\left(\mathbf{A}^{t-1} - \mathbf{I}\right)\mu\mathbf{U}_F^T \mathbf{D_S} \mathbf{w}(t) \\ &= -\frac{1}{\mu}\left(\mathbf{U}_F^T \mathbf{D_S} \mathbf{U}_F\right)^{-1}\mu\left(\mathbf{A}^{t-1} - \mathbf{I}\right)\mathbf{U}_F^T \mathbf{D_S} \mathbf{w}(t) \\ &= -\left(\mathbf{U}_F^T \mathbf{D_S} \mathbf{U}_F\right)^{-1}\left(\mathbf{A}^{t-1} - \mathbf{I}\right)\mathbf{U}_F^T \mathbf{D_S} \mathbf{w}(t)\end{aligned} \tag{15}$$

Based on (12) and (16), it can be obtained that

$$\Delta\hat{\mathbf{s}}_F(t) = -\mathbf{A}^{t-1}\mathbf{s}_F - \left(\mathbf{U}_F^T \mathbf{D_S} \mathbf{U}_F\right)^{-1}\left(\mathbf{A}^{t-1} - \mathbf{I}\right)\mathbf{U}_F^T \mathbf{D_S} \mathbf{w}(t) \tag{16}$$

Since $\mathrm{E}\left[\mathbf{w}(t)\mathbf{w}^T(t)\right] = \mathbf{C}_w(t) = \mathbf{C}_w$, according to (6) and (16), the $\mathrm{MSD}_G(t)$ of the GSP LMS algorithm can be obtained as follows.

$$\begin{aligned}&\mathrm{MSD}_G(t) \\ &= \Delta\hat{\mathbf{s}}_F^T(t)\Delta\hat{\mathbf{s}}_F(t) \\ &= \left[\mathbf{s}_F^T\left(\mathbf{A}^{t-1}\right)^T + \mathbf{w}^T(t)\mathbf{D_S}\mathbf{U}_F\left(\mathbf{A}^{t-1} - \mathbf{I}\right)^T\left(\left(\mathbf{U}_F^T \mathbf{D_S} \mathbf{U}_F\right)^{-1}\right)^T\right] \\ &\quad \times \left[\mathbf{A}^{t-1}\mathbf{s}_F - \left(\mathbf{U}_F^T \mathbf{D_S} \mathbf{U}_F\right)^{-1}\left(\mathbf{A}^{t-1} - \mathbf{I}\right)\mathbf{U}_F^T \mathbf{D_S} \mathbf{w}(t)\right] \\ &= \left\|\mathbf{U}_F \mathbf{A}^{t-1}\mathbf{s}_F\right\|_2^2 \\ &\quad + 2\mathrm{tr}\left[\mathbf{U}_F\left(\mathbf{U}_F^T \mathbf{D_S} \mathbf{U}_F\right)^{-1}\left(\mathbf{A}^{t-1} - \mathbf{I}\right)\mathbf{U}_F^T \mathbf{D_S} \mathrm{diag}\left(\mathrm{sqrt}(\mathbf{C}_w)\right)\mathbf{s}_F^T\left(\mathbf{A}^{t-1}\right)^T \mathbf{U}_F^T\right] \\ &\quad + \left\|\mathbf{U}_F\left(\mathbf{U}_F^T \mathbf{D_S} \mathbf{U}_F\right)^{-1}\left(\mathbf{A}^{t-1} - \mathbf{I}\right)\mathbf{U}_F^T \mathbf{D_S} \mathrm{diag}\left(\mathrm{sqrt}(\mathbf{C}_w)\right)\right\|_2^2\end{aligned}$$
$$\tag{17}$$

## B. Transient Error Analysis of the GSP RLS Algorithm

According to (4) and (8), it can be concluded that

$$\Delta \hat{\mathbf{s}}_F(t+1) = \lambda \Delta \hat{\mathbf{s}}_F(t) + (1-\lambda)\mathbf{M'U}_F^T \mathbf{D_S} \mathbf{C}_w^{-1} \mathbf{w}(t) \quad (18)$$

By the same definition $\Delta \hat{\mathbf{s}}_F(1) = -\mathbf{s}_F$, $\Delta \hat{\mathbf{s}}_F(2)$ can be obtained as follows

$$\begin{aligned}\Delta \hat{\mathbf{s}}_F(2) &= \lambda \Delta \hat{\mathbf{s}}_F(1) + (1-\lambda)\mathbf{M'U}_F^T \mathbf{D_S} \mathbf{C}_w^{-1} \mathbf{w}(t) \\ &= -\lambda \mathbf{s}_F + (1-\lambda)\mathbf{M'U}_F^T \mathbf{D_S} \mathbf{C}_w^{-1} \mathbf{w}(t)\end{aligned} \quad (19)$$

The derivation of $\Delta \hat{\mathbf{s}}_F(3)$ is shown below

$$\begin{aligned}\Delta \hat{\mathbf{s}}_F(3) &= \lambda \Delta \hat{\mathbf{s}}_F(2) + (1-\lambda)\mathbf{M'U}_F^T \mathbf{D_S} \mathbf{C}_w^{-1} \mathbf{w}(t) \\ &= \lambda \left[-\lambda \mathbf{s}_F + (1-\lambda)\mathbf{M'U}_F^T \mathbf{D_S} \mathbf{C}_w^{-1} \mathbf{w}(t)\right] \\ &\quad + (1-\lambda)\mathbf{M'U}_F^T \mathbf{D_S} \mathbf{C}_w^{-1} \mathbf{w}(t) \\ &= -\lambda^2 \mathbf{s}_F + \left[\lambda - \lambda^2 + 1 - \lambda\right]\mathbf{M'U}_F^T \mathbf{D_S} \mathbf{C}_w^{-1} \mathbf{w}(t) \\ &= -\lambda^2 \mathbf{s}_F + (1-\lambda^2)\mathbf{M'U}_F^T \mathbf{D_S} \mathbf{C}_w^{-1} \mathbf{w}(t)\end{aligned} \quad (20)$$

According to (18), (19) and (20), combined with mathematical induction, it can be drawn that

$$\Delta \hat{\mathbf{s}}_F(t) = -\lambda^{t-1}\mathbf{s}_F + (1-\lambda^{t-1})\mathbf{M'U}_F^T \mathbf{D_S} \mathbf{C}_w^{-1} \mathbf{w}(t) \quad (21)$$

The noise term $\mathbf{w}(t)$ is unavailable in practice, and since $E[\mathbf{w}(t)\mathbf{w}^T(t)] = \mathbf{C}_w(t)$, i.e., $\mathbf{C}_w(t)$ is the covariance matrix of $\mathbf{w}(t)$, thus, according to (6) and (21), the $\mathrm{MSD}_G(t)$ of the GSP RLS algorithm is summarized as follows

$$\begin{aligned}&\mathrm{MSD}_G(t) \\ &= \Delta \hat{\mathbf{s}}_F^T(t) \Delta \hat{\mathbf{s}}_F(t) \\ &= \left[-\lambda^{t-1}\mathbf{s}_F^T + (1-\lambda^{t-1})\mathbf{w}^T(t)(\mathbf{C}_w^{-1})^T \mathbf{D_S} \mathbf{U}_F (\mathbf{M'})^T\right] \\ &\quad \times \left[-\lambda^{t-1}\mathbf{s}_F + (1-\lambda^{t-1})\mathbf{M'U}_F^T \mathbf{D_S} \mathbf{C}_w^{-1} \mathbf{w}(t)\right] \\ &= \lambda^{2t-2}\|\mathbf{s}_F\|_2^2 \\ &\quad + 2(\lambda^{t-1}-1)\lambda^{t-1}\mathrm{tr}\left[\mathbf{s}_F^T\left(\mathbf{U}_F^T \mathbf{D_S} \mathbf{C}_w^{-1} \mathbf{D_S} \mathbf{U}_F\right)^{-1}\mathbf{U}_F^T \mathbf{D_S} \mathbf{C}_w^{-1} \mathrm{diag}(\mathrm{sqrt}(\mathbf{C}_w))\right] \\ &\quad + (\lambda^{t-1}-1)^2 \left\|\left(\mathbf{U}_F^T \mathbf{D_S} \mathbf{C}_w^{-1} \mathbf{D_S} \mathbf{U}_F\right)^{-1}\mathbf{U}_F^T \mathbf{D_S} \mathbf{C}_w^{-1} \mathrm{diag}(\mathrm{sqrt}(\mathbf{C}_w))\right\|_2^2\end{aligned} \quad (22)$$

## IV. EXPERIMENTAL VERIFICATION

The July temperature data of the Brazilian temperature dataset is used to validate the theoretical transient errors analysis of the GSP LMS and GSP RLS algorithm [24], [25]. The July data of the Brazilian temperature dataset was collected by using 299 weather stations, that is $N = 299$. Each station is connected to its $K$ closest stations in the neighborhood.

Based on the above temperature dataset, two graph-structured scenarios are created, which are shown as follows:
Case I: $K = 8$, $F = 200$, $|\mathbf{S}| = 210$.
Case II: $K = 16$, $F = 160$, $|\mathbf{S}| = 210$.
Case I and Case II are corresponding to Fig. 1 and Fig. 2, respectively.

When $F$ and $|\mathbf{S}|$ are determined, $\mathbf{U}_F$ and $\mathbf{D_S}$ are determined accordingly. The evaluation parameter $\mathrm{MSD}_G(t)$ is as follows.

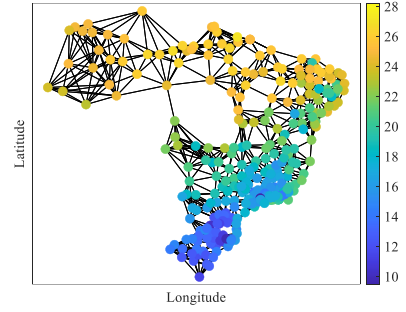

Fig. 1. Graph based on July data in Brazil dataset(Case I)

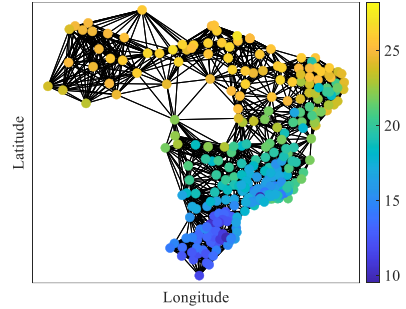

Fig. 2. Graph based on July data in Brazil dataset(Case II)

$$\mathrm{MSD}_G(t) = 10 \times \log_{10} \|\hat{\mathbf{x}}_o(t) - \mathbf{x}_o\|_2^2 \quad (23)$$

$\mathbf{C}_w$ is used to produce the noise signal

$$\mathbf{C}_w = \mathrm{diag}(N_a \mathbf{a} + N_b \mathbf{b}) \quad (24)$$

where $\mathbf{a}$ is a vector whose components satisfy a Gaussian distribution, $\mathbf{b}$ is a vector of the same shape as $\mathbf{a}$ with all components 1. $N_a$, $N_b$ are the coefficients of $\mathbf{a}$ and $\mathbf{b}$, respectively.

The following three Gaussian noise scenarios are set up for simulation verification.

$$N_a = 0.012, \ N_b = 0 \quad (i)$$

$$N_a = 0.05, \ N_b = 0 \quad (ii)$$

$$N_a = 0.05, \ N_b = 0.05 \quad (iii)$$

All the results of the following experiments are obtained by taking the average of 50 independent experiments.

### A. Simulation Verification of Transient Errors of the GSP LMS Algorithm

In this section, based on the three set up Gaussian noise scenarios and the two different graph structures, i.e., Case I, Case II, the experimental validation of the theoretical transient error analysis of the GSP LMS algorithm is carried out, as shown in Fig. 3. The number of iterations is set to 1000. The step size of the GSP LMS algorithm in Case I is set to 0.43 and 1.57, respectively. And the step size of the GSP LMS algorithm in Case II is set to 0.2 and 1.1, respectively. From shown in Fig. 3, it can be found that the theoretical analysis of the transient error of the GSP LMS algorithm proposed in this paper and the actual simulation results of the GSP LMS algorithm achieve a better match in different noise cases and different graph adjacency structures.

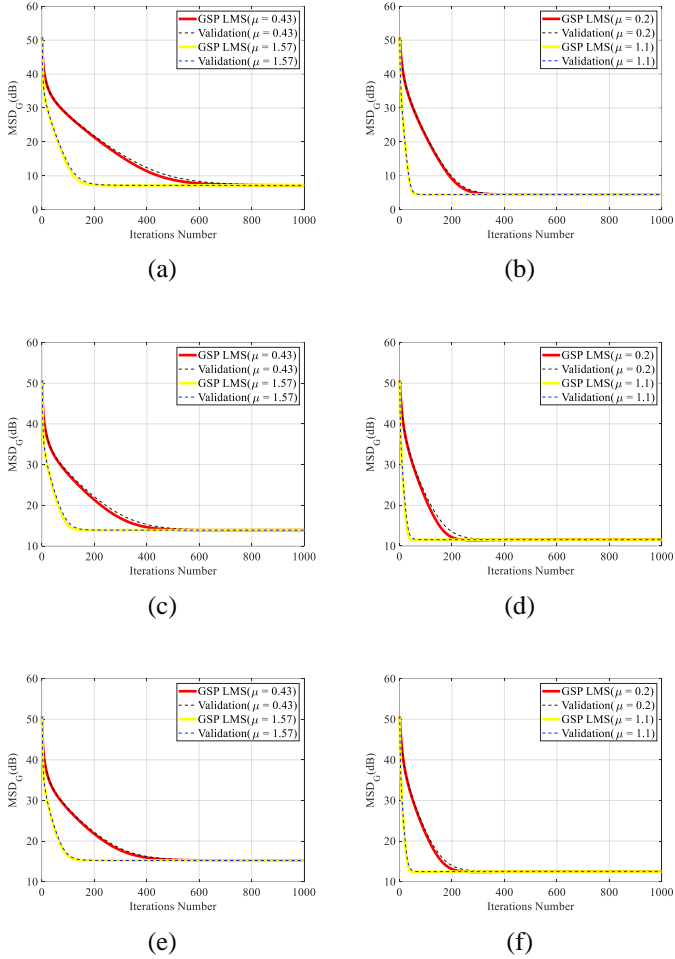

**Fig. 3.** Simulation verification of transient errors analysis of the GSP LMS algorithm. (a) $\mathbf{C}_w(\text{i})$ under case I (b) $\mathbf{C}_w(\text{i})$ under case II (c) $\mathbf{C}_w(\text{ii})$ under case I (d) $\mathbf{C}_w(\text{ii})$ under case II (e) $\mathbf{C}_w(\text{iii})$ under case I (f) $\mathbf{C}_w(\text{iii})$ under case II

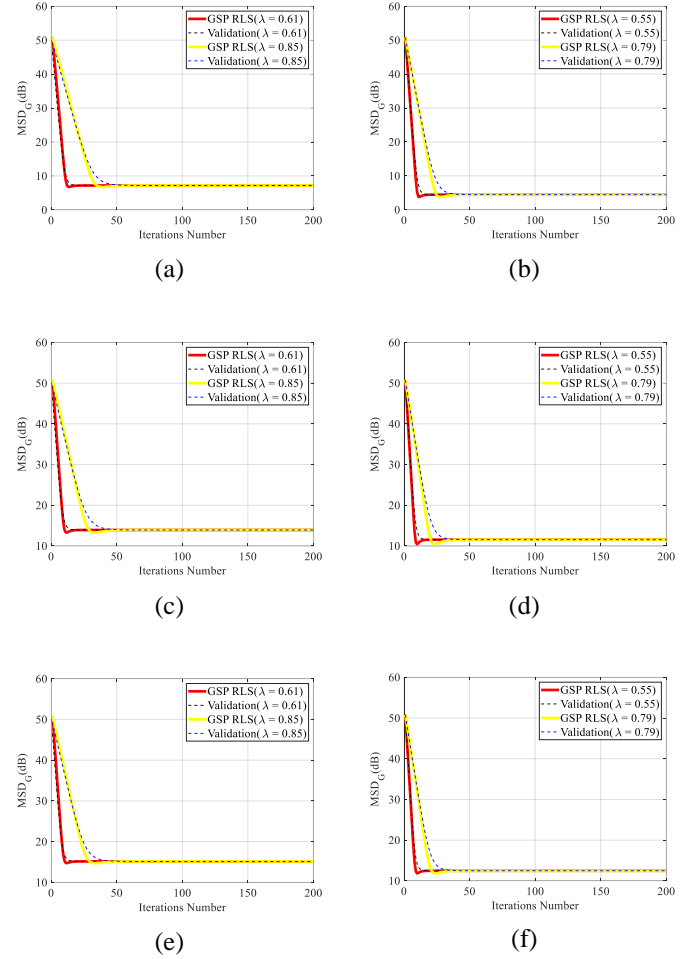

**Fig. 4.** Simulation verification of transient errors analysis of the GSP RLS algorithm. (a) $\mathbf{C}_w(\text{i})$ under case I (b) $\mathbf{C}_w(\text{i})$ under case II (c) $\mathbf{C}_w(\text{ii})$ under case I (d) $\mathbf{C}_w(\text{ii})$ under case II (e) $\mathbf{C}_w(\text{iii})$ under case I (f) $\mathbf{C}_w(\text{iii})$ under case II

### B. Simulation Verification of Transient Errors of the GSP RLS Algorithm

Based on the relevant noise environment setups in (24), (i), (ii), (iii), the two graph structures shown in Fig. 1 and Fig. 2, the validation of the theoretical transient error expressions of the GSP RLS algorithm is performed in this section. as shown in Fig. 4. The number of iterations is set to 200. The forgetting factor $\lambda$ of the GSP RLS algorithm in Case I is set to 0.61 and 0.85, respectively. And forgetting factor $\lambda$ of the GSP RLS algorithm in Case II is set to 0.55 and 0.79, respectively. From shown in Fig. 4, it can be found that the theoretical analysis of the transient error of the GSP RLS algorithm proposed in this paper and the actual simulation results of the GSP LMS algorithm achieve a better match in different noise cases and different graph adjacency structures.

### V. CONCLUSION

In this paper, the theoretical transient errors analysis of the GSP LMS algorithm and GSP RLS algorithm is carried out. Based on the respective frequency domain error expressions and combining mathematical induction, the transient error expressions of the GSP LMS algorithm and GSP RLS algorithm is obtained, respectively. Based on the July temperature data from the Brazilian temperature dataset, the theoretical transient errors analysis of the GSP LMS algorithm and GSP RLS algorithm under different Gaussian noise environments and different graph adjacency matrices is validated in this paper, respectively. It should be noted that the current work is only limited to the case where the spectrally sparse parameter $F$ and the sampling set are priori and fixed. We will further investigate the transient error behavior of the related algorithms in the case of time-varying $F$ and sampling set in the future.